\documentclass[12pt,preprint]{aastex}

\usepackage{graphicx}
\usepackage{txfonts}
\usepackage{longtable}
\usepackage{lscape}
\usepackage{natbib}

\shorttitle{The Enigmatic Polarization of the D$_1$ Lines of Ba~{\sc ii} and 
Na~{\sc i}}
\shortauthors{Belluzzi and Trujillo Bueno}

\begin{document}

\title{A key physical mechanism for understanding the enigmatic \\ 
linear polarization of the solar Ba~{\sc ii} and Na~{\sc i} D$_1$ lines}

\author{{\sc Luca Belluzzi}\altaffilmark{1,2} {\sc and Javier Trujillo 
Bueno}\altaffilmark{1,2,3}}
\altaffiltext{1}{Instituto de Astrof\'isica de Canarias, E-38205 La Laguna, 
Tenerife, Spain}
\altaffiltext{2}{Departamento de Astrof\'isica, Facultad de F\'isica, 
Universidad de La Laguna, E-38206 La Laguna, Tenerife, Spain}
\altaffiltext{3}{Consejo Superior de Investigaciones Cient\'ificas, Spain}

\begin{abstract}
The linearly polarized spectrum of the solar limb radiation produced by 
scattering processes is of great diagnostic potential for exploring the 
magnetism of the solar atmosphere.
This spectrum shows an impressive richness of spectral details and enigmatic 
$Q/I$ signals, whose physical origin must be clearly understood before they 
can be exploited for diagnostic purposes.
The most enduring enigma is represented by the polarization signals
observed in the D$_1$ resonance lines of Na~{\sc i} (5896~\AA) and Ba~{\sc ii} 
(4934~\AA), which were expected to be intrinsically unpolarizable.
The totality of sodium and 18\% of barium have hyperfine structure (HFS), and 
it has been argued that the only way to produce a scattering polarization 
signal in such lines is through the presence of a substantial amount of atomic 
polarization in their lower HFS levels.
The strong sensitivity of these long-lived levels to depolarizing mechanisms 
led to the paradoxical conclusion that the observed D$_1$-line polarization is 
incompatible with the presence in the lower solar chromosphere of inclined 
magnetic fields sensibly stronger than 0.01 G.
Here we show that by properly taking into account the fact that the solar 
D$_1$-line radiation has a non-negligible spectral structure over the short 
frequency interval spanned by the HFS transitions, it is possible to produce 
scattering polarization signals in the D$_1$ lines of Na~{\sc i} and 
Ba~{\sc ii} without the need of ground-level polarization. 
The resulting linear polarization is not so easily destroyed by elastic 
collisions and/or magnetic fields.
\end{abstract}


\section{Introduction}
When observed with high-sensitivity spectropolarimeters, the linearly polarized 
spectrum of the solar radiation coming from quiet regions close to the limb 
(the so-called ``second solar spectrum'') reveals a complex structure and an 
impressive richness of spectral details \citep{Ste97}.
We know that the second solar spectrum is the observational signature of atomic 
polarization (i.e., population imbalances and quantum interference between 
different magnetic sublevels), produced by the absorption and scattering of 
anisotropic radiation.
However, the physics of scattering polarization in an optically thick plasma, 
such as the extended solar atmosphere, is very complicated and a general 
theory for multilevel systems is still lacking.
It is thus natural that many of the $Q/I$ signals of the second solar spectrum
were considered as enigmatic by their discoverers, as well as a true challenge 
for the theorists \citep[see the review by][]{JTB09}.\footnote{The second solar 
spectrum is generally expressed through the ratio between the Stokes parameters 
$Q$ and $I$, with the reference direction for positive $Q$ parallel to the 
nearest solar limb.}
At present, the most challenging $Q/I$ signals are those observed in the D$_1$ 
lines of Na~{\sc i} and Ba~{\sc ii} at 5896~{\AA} and 4934~{\AA}, respectively 
\citep[see Figures 2 and 5 of][]{Ste97}. 
The aim of this Letter is to point out a physical mechanism that might be of 
key importance for explaining such enigmatic linear polarization profiles.

The polarization signals observed by \citet{Ste97} and \citet{Ste00} in the 
D$_1$ lines of Na~{\sc i} and Ba~{\sc ii} were considered enigmatic because 
these lines were expected to be intrinsically unpolarizable.
This is because they result from transitions between an upper and a lower level 
with $J=1/2$ ($J$ being the total angular momentum).
Such levels cannot carry atomic alignment, so that no linear polarization can 
in principle be produced in these lines in weakly magnetized regions of the 
solar atmosphere.\footnote{An atomic level is said to be aligned when sublevels 
with different values of $|M|$ ($M$ being the magnetic quantum number) are 
unevenly populated. When the upper and/or lower level of a line transition is 
aligned, then the emergent radiation is in general linearly polarized, even in 
the absence of a magnetic field \citep[e.g.,][]{Lan04}.}
However, 100\% of sodium and about 18\% of barium have hyperfine structure 
(HFS). Both sodium ($^{23}$Na) and the barium isotopes with HFS ($^{135}$Ba and 
$^{137}$Ba) have nuclear spin $I=3/2$. 
In such isotopes, the upper and lower $J$-levels of the D$_1$ line transition 
thus split into two HFS $F$-levels ($F=1$ and $F=2$, with $F$ the total, 
electronic plus nuclear, angular momentum), which can carry atomic alignment.
In fact, taking the HFS of sodium into account, \citet{Lan98} concluded 
that if there is a substantial amount of atomic alignment in the ground level 
(i.e., in the lower $F=1$ and $F=2$ levels of the sodium D$_1$ and D$_2$ 
lines), then a significant $Q/I$ signal is produced in the D$_1$ line. 
In his investigation of the sodium doublet, \citet{Lan98} considered the 
unmagnetized case and assumed frequency-coherent scattering and a constant 
pumping radiation field within each D-line.

In Landi Degl'Innocenti's (1998) modeling of the sodium D$_1$ line, the ensuing 
$Q/I$ profile vanishes if the alignment of the ground $F$-levels is destroyed.
Such lower $F$-levels are long-lived and therefore their atomic alignment is 
very sensitive to depolarizing mechanisms, such as elastic collisions with 
neutral hydrogen atoms and/or the Hanle effect of very weak (mG) non-vertical 
magnetic fields. 
Accordingly, \citet{Lan98} concluded that the $Q/I$ signal detected by 
\citet{Ste97} in the sodium D$_1$ line implies that depolarization does not 
occur in the lower solar chromosphere (where the line-core of the sodium 
D-lines originates), which would seem to rule out the existence of tangled 
magnetic fields and of inclined, canopy-like fields stronger than 0.01~G. 
He also pointed out that this is difficult to understand because there is 
substantial evidence from other types of observations for both types of 
magnetic field. 
This paradox was later reinforced by detailed theoretical investigations on 
the magnetic sensitivity of the D-lines of Na~{\sc i} \citep{JTB02,Cas02} and 
of Ba~{\sc ii} \citep{Bel07}, as well as on the depolarizing role of elastic 
collisions with neutral hydrogen atoms \citep{Ker02}.
It is also noteworthy that the scattering polarization profile obtained by 
\citet{Lan98} is antisymmetric, in contrast with the symmetric peak observed 
by \citet{Ste97}, but in agreement with the observations of other researchers 
\citep[see][for a review]{JTB09}.
 
The physical mechanism discussed in this Letter does not require the presence 
of atomic polarization in the lower level.
Indeed, we will show that by properly taking into account HFS and partial 
frequency redistribution (PRD) effects, even without any atomic polarization 
in the (long-lived) lower $F$-levels, the scattering of anisotropic D$_1$-line 
radiation in the solar atmosphere can produce significant linear polarization 
in the D$_1$ lines of Ba~{\sc ii} and Na~{\sc i}. 

\section{The physical problem}
We model the intensity and linear polarization profiles of the D$_1$ lines of 
Ba~{\sc ii} and Na~{\sc i} by solving the full non-LTE radiative transfer 
problem for polarized radiation in one-dimensional semi-empirical models of the 
solar atmosphere, taking PRD effects into account.
To this aim, we apply the redistribution matrix approach outlined in 
\citet{Bel12}, which can be applied also to the case of a two-level atom with 
HFS, under the assumptions that the lower $F$-levels are unpolarized and 
infinitely sharp.
As previously discussed, these approximations appear to be suitable since the 
lower level of the D$_1$ lines of Ba~{\sc ii} and Na~{\sc i} is the 
(long-lived) ground level.
We recall that our redistribution matrix is given by the linear combination of 
two terms, which describe purely coherent scattering ($R_{\rm II}$) and 
completely redistributed scattering ($R_{\rm III}$) in the atom rest frame, 
respectively.

We investigate the D$_1$ line of barium accounting for the contribution of all 
its seven stable isotopes. The two odd isotopes ($^{135}$Ba and $^{137}$Ba, 
both with nuclear spin $I=3/2$) are described through a two-level model atom 
with HFS, the five even isotopes (about 82.18\% in abundance) are described 
through a two-level model atom without HFS. 
The D$_1$ line of sodium is modeled through a two-level model atom with HFS 
(sodium has a single stable isotope, $^{23}$Na, with $I=3/2$).

As clear from the Grotrian diagram shown in the left panel of Figure~1, in 
the isotopes with HFS, each upper $F$-level can be excited from the two lower 
$F$-levels through two transitions that are very close in frequency (about 
70~m{\AA} apart in the barium isotopes with HFS, and about 20~m{\AA} apart in 
sodium).
Due to such small frequency separation, the assumption that the pumping 
radiation field is the same in the two HFS transitions appears to be an 
excellent approximation for modeling these lines.
However, under this hypothesis, and assuming that the lower level is 
unpolarized, no atomic polarization can be produced in the upper $F$-levels, 
and therefore no polarization can be obtained in the emergent radiation.
This can be seen, for example, from the statistical equilibrium equations 
for a two-level atom with HFS derived in Landi Degl'Innocenti \& Landolfi 
(2004, hereafter LL04).
Under the hypothesis of unpolarized lower level, such equations can be 
analytically solved, and the multipole components of the density matrix 
of the upper levels result to be given by
\begin{eqnarray}
	\rho^K_Q(F_u, F_u^{\prime}) \!\! & =  & \!\! 
	\frac{\mathbb{N}_{\ell}}{\mathcal{N}} \,
	\frac{\sqrt{3 (2F_u + 1) (2F_u^{\prime} + 1)}}{2I + 1} \,
	\frac{B(J_{\ell} \rightarrow J_u)}{A(J_u \rightarrow J_{\ell}) + 
	2 \pi {\rm i} \nu_{F_u, F_u^{\prime}}} \, \sum_{F_{\ell}} \,
	(-1)^{1 + F_{\ell} + F_u + Q} \nonumber \\
	& & \times \, (2F_{\ell} + 1) \, 
	\left\{ 
	\begin{array}{c c c}
		1 & 1 & K \\
		F_u & F^{\prime}_u & F_{\ell}
	\end{array}
	\right\}
	\left\{
	\begin{array}{c c c}
		J_u & J_{\ell} & 1 \\
		F_{\ell} & F_u & I
	\end{array}
	\right\}
	\left\{
	\begin{array}{c c c}
		J_u & J_{\ell} & 1 \\
		F_{\!\ell} & F^{\prime}_{\!u} & I
	\end{array}
	\right\} 
	J^K_{-Q}(\nu_0) \; ,
\end{eqnarray}
where $A(J_u \rightarrow J_{\ell})$ and $B(J_{\ell} \rightarrow J_u)$ are 
the Einstein coefficients for spontaneous emission and for absorption, 
respectively, $\mathcal{N}$ is the total number density of atoms, 
$\mathbb{N}_{\ell}$ is the number density of atoms in the lower $J$-level, 
and $\nu_{F_u, F_u^{\prime}}$ is the frequency separation between the HFS 
levels $F_u$ and $F_u^{\prime}$.
Since the incident field (described in Equation~(1) through the radiation field 
tensor $J^K_Q(\nu_0)$, see Equation~(5.157) of LL04) is assumed to be constant 
across the various HFS transitions (this is one of the hypotheses the theory 
described in LL04 is based on), it is possible to perform the sum over 
$F_{\ell}$, thus obtaining
\begin{eqnarray}
	\rho^K_Q(F_u, F_u^{\prime}) \!\! & =  & \!\! 
	\frac{\mathbb{N}_{\ell}}{\mathcal{N}} \,
	\frac{\sqrt{3 (2F_u + 1) (2F_u^{\prime} + 1)}}{2I + 1} \,
	\frac{B(J_{\ell} \rightarrow J_u)}{A(J_u \rightarrow J_{\ell}) + 
	2 \pi {\rm i} \nu_{F_u, F_u^{\prime}}} \,
	(-1)^{1 - J_{\ell} + I + F^{\prime}_u + K + Q} \nonumber \\
	& & \times \, 
	\left\{ 
	\begin{array}{c c c}
		1 & 1 & K \\
		J_u & J_u & J_{\ell}
	\end{array}
	\right\}
	\left\{
	\begin{array}{c c c}
		J_u & J_u & K \\
		F_u & F^{\prime}_u & I
	\end{array}
	\right\}
	J^K_{-Q}(\nu_0) \; .
\end{eqnarray}
Observing that the first 6-$j$ symbol in the second line of Equation~(2) 
vanishes for $J_u = 1/2$ and $K=2$, we immediately see that under the 
above-mentioned hypotheses, no atomic polarization (either alignment or 
$F$-state interference, both being described by the multipole components with 
$K=2$) can be induced in the upper $F$-levels of the D$_1$ line transition.
Indeed, in the previous investigations on this line, in which the pumping 
radiation field was assumed to be identical in the various HFS transitions, 
either because the investigation was carried out by applying the complete 
frequency redistribution (CRD) theory presented in LL04 \citep[see][]{JTB02}, 
or because this was assumed by definition, though considering coherent 
scattering \citep[see][]{Lan98}, the authors concluded that the only way to 
obtain a scattering polarization signal in the core of the D$_1$ line is by 
assuming that a given amount of atomic polarization is present in the lower 
level.

In this investigation, on the contrary, we neglect lower level polarization but 
we take into account, through the $R_{\rm II}$ term of the redistribution 
matrix, that the pumping radiation field can, in principle, be different in the 
various HFS transitions (see the right panel of Figure~1 for the case of 
barium).
Although such difference is in many cases very small, we find that this allows 
atomic polarization to be induced in the upper $F$-levels, giving rise to an 
appreciable polarization signal in the core of the D$_1$ line (see the next 
section). 
According to this physical mechanism, the amplitude of the linear polarization 
signal produced in the D$_1$ line is expected to be higher when the HFS 
splitting of the lower level is larger.

The two-level atom models (either with or without HFS) considered in this
investigation do not allow us to take quantum interference between the upper 
$J$-levels of the D$_1$ and D$_2$ lines into account.
This physical ingredient can actually be safely neglected for the investigation 
of the Ba~{\sc ii} D$_1$ line, since it is about 380~{\AA} away from D$_2$, 
but it is known to play an important role in the modeling of the Na~{\sc i}
D-lines, which are only 6~{\AA} apart.
On the other hand, an accurate modeling of the scattering polarization in the 
Ba~{\sc ii} D$_1$ and D$_2$ lines would require to account for the radiative 
and collisional coupling of the upper levels of such lines with the metastable 
$^2{\rm D}_{3/2}$ and $^2{\rm D}_{5/2}$ levels.
This coupling, which is not present in neutral sodium since it does not have 
such metastable levels, cannot be accounted for within our two-level atom 
modeling.
As we shall see below, in spite of such limitations, our approach is suitable 
to show that the physical mechanism identified in this investigation might be 
of key relevance for understanding the enigmatic D$_1$ line polarization.

\section{Results}   
Figure~2 shows the emergent Stokes $I$ and $Q/I$ profiles of the Ba~{\sc ii} 
D$_1$ line calculated in the semi-empirical model C of Fontenla et al. 
(1993; hereafter, FAL-C model). 

The profiles in the left panel of Figure~2 have been calculated in the limit of 
CRD, following the theoretical approach of LL04. 
As discussed in Section~2, within the framework of such theory, the absorption 
of radiation from an unpolarized lower level cannot induce any atomic 
polarization in the upper $F$-levels and therefore the D$_1$ line must behave 
as intrinsically unpolarizable. 
Indeed, as shown in the lower left panel of Figure~2, the D$_1$ line only 
produces a depolarization of the continuum polarization level (our calculations 
account for the contribution of a coherent polarized continuum). 

The right panels of Figure~2 show the results of PRD calculations. 
In this case, through the $R_{\rm II}$ term of the redistribution matrix, we 
account for the exact frequency dependence of the pumping radiation field 
within the D$_1$ line.
As expected, no signal is obtained if only $^{138}$Ba (which is devoid of HFS) 
is considered (see the dotted line).
However, if the contribution of the small fraction ($\sim 18$\%) of the barium 
isotopes with HFS is taken into account, a conspicuous polarization signal is 
obtained for a line-of-sight (LOS) with $\mu = 0.1$ ($\mu$ being the cosine of 
the heliocentric angle).
The $Q/I$ signal shows a large positive peak (about 0.45\%) slightly 
blueshifted with respect to the line center, and a smaller negative peak 
(about $-0.15$\%) slightly redshifted.
Such $Q/I$ peaks decrease, though remaining significant, for larger $\mu$ 
values. 

As can be noted by comparing the top panels of Figure~2, the Stokes $I$ profile 
is not very sensitive to PRD effects, except at the very line center where the 
PRD calculation gives slightly smaller intensity values.
Particularly interesting, on the other hand, is the significant broadening of 
the Stokes $I$ profile produced by the small fraction of barium isotopes with 
HFS (compare the solid and dotted lines).

The results of Figure~3 are also noteworthy. 
The bottom panel shows the $Q/I$ profiles calculated in three different
semi-empirical models of the solar atmosphere (the temperature structure 
of such models is plotted in the upper panel). 
As can be observed, the $Q/I$ signals are not very different. 
This small sensitivity to the thermal model is good news, because the $Q/I$ 
signal of the Ba~{\sc ii} D$_1$ line produced by the physical mechanism 
discussed in Section~2 is sensitive to the upper-level Hanle effect and 
therefore to magnetic fields in the gauss range.

In Figure~4, we show the results that we have obtained for the D$_1$ line of 
Na~{\sc i}.
The sensitivity of the $Q/I$ signal of this stronger line to the thermal model 
is more noticeable, with the (relatively cool) FAL-X model producing the 
largest peaks.
Such $Q/I$ peaks are much weaker than those observed by \citet{Ste97} and than
those obtained in the Ba~{\sc ii} D$_1$ line.
The calculated $Q/I$ signals are much weaker in sodium than in barium because 
the HFS splitting of the ground level is about four times smaller in 
Na~{\sc i}.

\section{Concluding comments}
Two physical mechanisms able to produce a scattering polarization signal in the 
Na~{\sc i} and Ba~{\sc ii} D$_1$ lines have been identified up to now. 
HFS is a fundamental ingredient in both of them.
Such physical mechanisms are not mutually exclusive, and are probably at work 
simultaneously in the solar atmosphere, especially in the case of the 
Na~{\sc i} D$_1$ line.

One mechanism is that pointed out by \citet{Lan98}, which requires the presence
of a substantial amount of ground-level polarization.
As clarified by \citet{JTB02}, this mechanism, which operates even when the 
pumping radiation is constant within the D lines \citep[see also][]{Cas02}, 
requires taking into account the absorption of anisotropic radiation in the 
D$_2$ line (whose upper level has $J=3/2$) and to assume that depolarizing 
mechanisms do not completely destroy the atomic alignment that 
{\em repopulation pumping} produces in the long-lived lower $F$-levels. 
Under such circumstances, the atomic alignment of the lower and upper 
$F$-levels of the D$_1$ line (and therefore its scattering polarization 
signal) are sensitive to the same (mG) magnetic field strengths. 

In this Letter, we have shown that the absorption and scattering of anisotropic 
D$_1$-line radiation within the solar atmosphere can directly produce linear 
polarization in the D$_1$ line, without any need for ground-level polarization. 
This is possible if one takes into account that the solar D$_1$-line 
radiation has a non-negligible spectral structure over the short frequency 
interval spanned by the HFS transitions. 
The fact that small variations of the pumping radiation field between very 
close HFS transitions is able to create a significant amount of atomic 
polarization in the upper $F$-levels of the D$_1$ lines is certainly an 
unexpected and important result.
The resulting $Q/I$ signals are not so easily destroyed by elastic collisions 
and/or magnetic fields.

The amplitude and shape of the calculated $Q/I$ profile of the Ba~{\sc ii} 
D$_1$ line are similar to the observed ones\footnote{Although the calculated 
amplitude is in reality larger than the observed one, they can be easily 
brought into reasonable agreement by taking into account the spectral smearing 
produced by the limited spectral resolution of the observation and/or 
depolarizing mechanisms.} \citep[see Figure~5 of][]{Ste97}. 
The main discrepancy concerns the slightly redshifted peak, which reaches a 
negative value in the theoretical profile, while it remains positive in the 
observation.
In our opinion, such discrepancy does not invalidate our conclusion that the 
physical mechanism we have discussed in this Letter may be a key ingredient 
for explaining the enigmatic linear polarization of the Ba~{\sc ii} D$_1$ line.
This is because our Ba~{\sc ii} model atom does not include the metastable 
$^2{\rm D}_{3/2}$ and $^2{\rm D}_{5/2}$ levels, which may have a significant 
impact on the D$_1$ line polarization \citep[see][concerning the Ca~{\sc ii} 
IR triplet]{Man03}.

The amplitudes of the $Q/I$ profiles observed in the D$_1$ lines of Na~{\sc i} 
and Ba~{\sc ii} are of the same order of magnitude, but the calculated $Q/I$ 
signal of the Na~{\sc i} D$_1$ line is one order of magnitude smaller.
This is due to fact that the HFS splitting of the ground level of sodium is 
about four times smaller than that of barium.
We believe that this discrepancy with the observation does not rule out the 
possibility that the physical mechanism described in this Letter may play 
a key role also for explaining the enigmatic linear polarization of the 
Na~{\sc i} D$_1$ line.
We plan to include in our modeling the quantum interference between the 
$F$-levels pertaining to the two upper $J$-levels of the sodium doublet. 
This additional physical ingredient will probably influence the shape of the 
$Q/I$ profile, even in the neighborhood of the sodium D$_1$ line core.
In addition to this, we believe that if there is a non-negligible amount of 
ground-level polarization, then the physical mechanism discussed in this Letter 
should induce a larger $Q/I$ signal.

\acknowledgements
Financial support by the Spanish Ministry of Economy and Competitiveness and 
the European FEDER Fund through project AYA2010-18029 (Solar Magnetism and 
Astrophysical Spectropolarimetry) is gratefully acknowledged.

\begin{figure}
\centering
\includegraphics[width=1.0\textwidth]{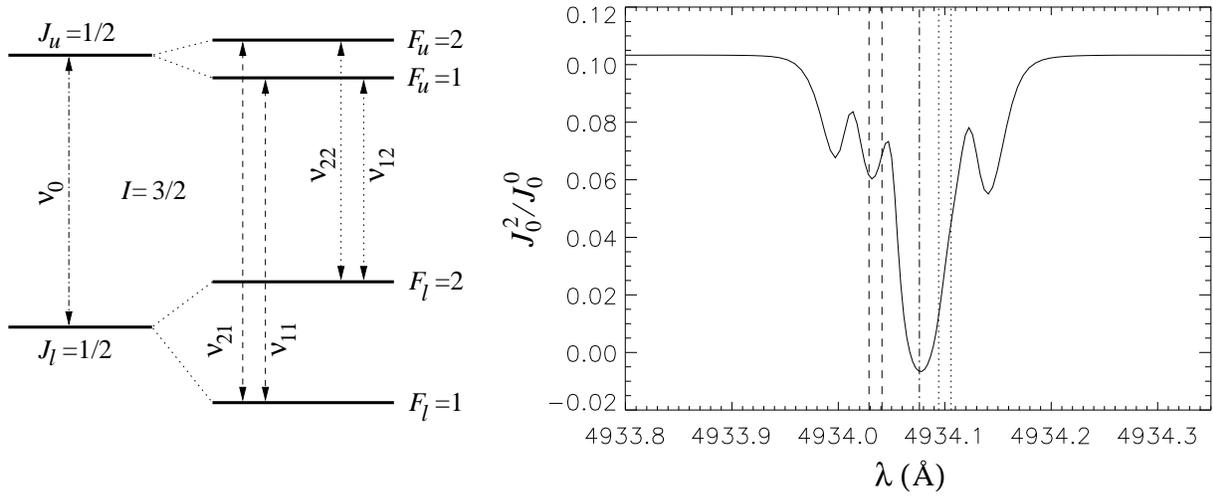}  
\caption{Left panel: Grotrian diagram showing the fine structure (FS) and HFS 
levels considered in our atomic models of sodium and of the barium isotopes 
with HFS (splittings are not drawn to scale). The FS and HFS components of 
the D$_1$ line transition are drawn on the diagram.
Right panel: anisotropy (expressed through the ratio between the $J^2_0$ and 
$J^0_0$ components of the radiation field tensor) of the solar radiation  
across the Ba~{\sc ii} D$_1$ line at 800~Km in the FAL-C model atmosphere 
(the height where the line center optical depth is unity for a LOS with 
$\mu=0.1$). 
The vertical lines indicate the wavelength positions of the four HFS components 
of $^{137}$Ba, and of the FS component of $^{138}$Ba.}
\label{fig:fig1}
\end{figure}

\begin{figure}
\centering
\includegraphics[width=1.0\textwidth]{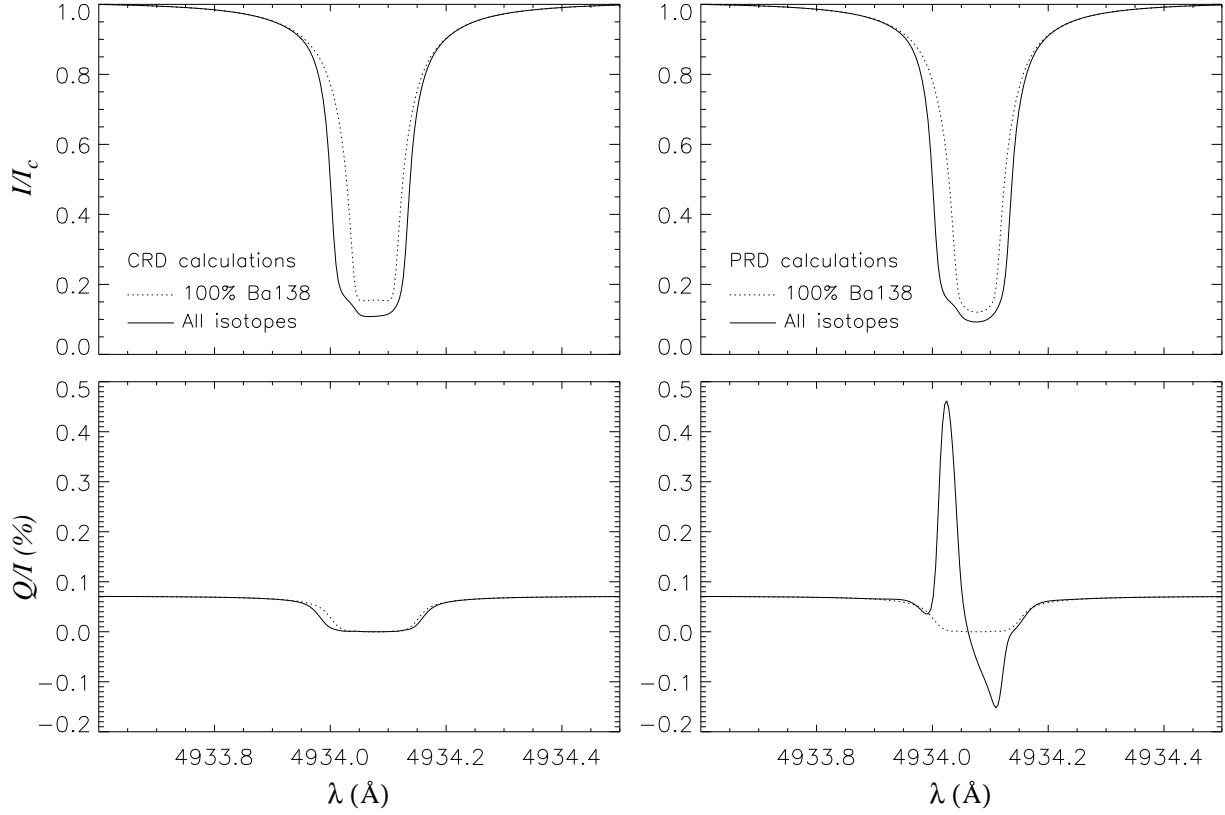}  
\caption{Left: intensity (upper panel) and $Q/I$ (lower panel) profiles of 
the Ba~{\sc ii} D$_1$ line, as calculated by applying the CRD theory presented 
in LL04, considering the FAL-C atmospheric model, and a LOS with $\mu=0.1$. 
The solid line represents the solution obtained by considering the contribution 
of all the seven isotopes of barium. 
The dotted line represents the solution obtained by assuming that 100\% of 
barium is $^{138}$Ba (which is devoid of HFS).
Right: same as left panels, but for PRD calculations.}
\label{fig:fig2}
\end{figure}

\begin{figure}
\centering
\includegraphics[width=0.7\textwidth]{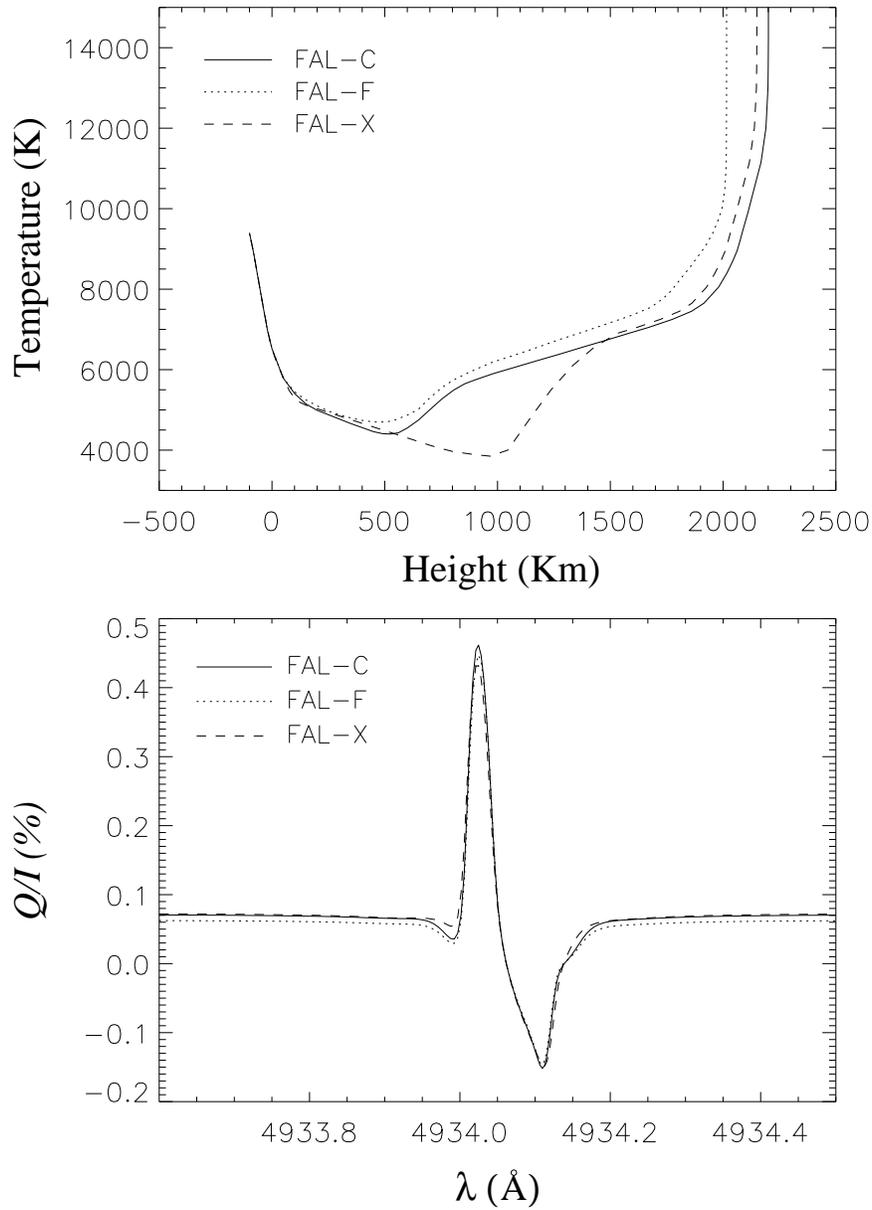}  
\caption{Upper panel: variation of the temperature with height in the 
semi-empirical models C and F of \citet{Fon93}, and in the semi-empirical model
M$_{\rm CO}$ (also known as FAL-X) of \citet{Avr95}.
Lower panel: PRD $Q/I$ profiles of the Ba~{\sc ii} D$_1$ line calculated in the 
same atmospheric models, for a LOS with $\mu=0.1$.}
\label{fig:fig3}
\end{figure}

\begin{figure}
\centering
\includegraphics[width=0.7\textwidth]{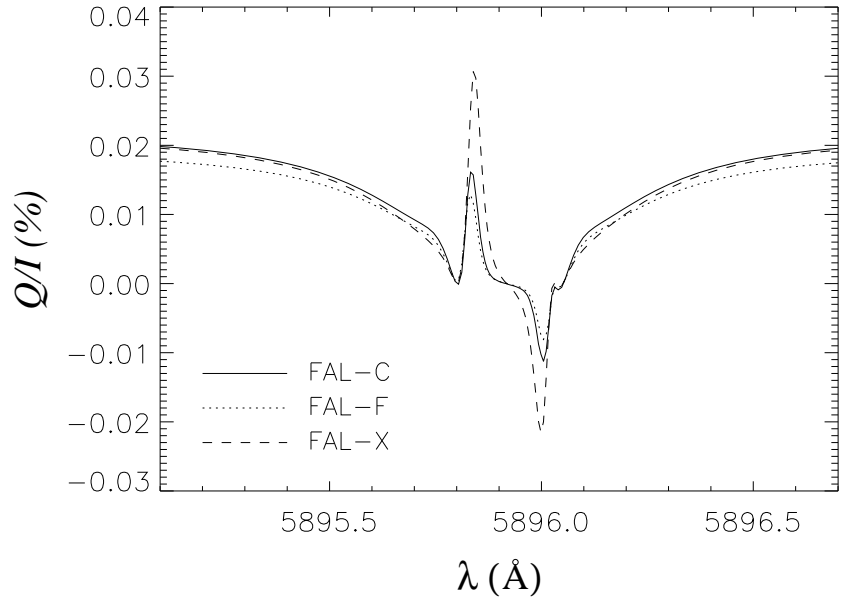}  
\caption{PRD $Q/I$ profiles of the Na~{\sc i} D$_1$ line calculated in the 
semi-empirical models C and F of \citet{Fon93}, and in the semi-empirical model 
M$_{\rm CO}$ (also known as FAL-X) of \citet{Avr95}, for a LOS with $\mu=0.1$.}
\label{fig:fig4}
\end{figure}

\end{document}